\documentclass[aps,prb,twocolumn,groupedaddress]{revtex4}

\usepackage{graphicx}
\usepackage{bm}

\begin{document}

%\preprint{}


\title{The influence of electrostatic potentials on
  the apparent $s$-$d$ exchange energy in III-V diluted magnetic semiconductors}



\author{C.\ \'Sliwa}
\email{sliwa@ifpan.edu.pl}
\affiliation{Institute of Physics, Polish Academy of Sciences, al.\ 
  Lotnik\'ow 32/46, 02-668 Warszawa, Poland}

\author{T.\ Dietl}
\email{dietl@ifpan.edu.pl}
%\thanks{}
\affiliation{Institute of Physics, Polish Academy of Sciences and ERATO 
Semiconductor Spintronics Project, Japan Science and Technology, 
al.\ Lotnik\'ow 32/46, 02-668 Warszawa, Poland}
\affiliation{Institute of Theoretical Physics, Warsaw University, 
00-681 Warszawa, Poland}


\date{\today}

\begin{abstract}
  The muffin-tin model of an effective-mass electron interacting with
  magnetic ions in semiconductors is extended to incorporate 
  electrostatic potentials that are present in the case of  Mn-based
  III-V compounds (${\rm Ga}_{1-x}{\rm Mn}_x {\rm N}$, ${\rm
  Ga}_{1-x}{\rm Mn}_x {\rm As}$). Since the conduction band 
  electron is repelled from negatively charged magnetic ions 
  and attracted by compensating donors, the \emph{apparent} value of 
  the $s$-$d$ exchange coupling $N_0 \alpha$ is reduced. It is shown 
  that the magnitude of this effect 
  increases when $x$ diminishes. Our
  model may explain an unusual behavior of electron spin splitting observed
  recently in those two materials in the Mn concentration range $x \le 0.2$\%.
\end{abstract}

%\keywords{}

\maketitle

Owing to the possibility of a gradual incorporation of
magnetism to the well-known semiconductor matrices, diluted magnetic 
semiconductors (DMS)\cite{Furd88,Diet94,Mats02} offer unprecedented opportunity for
examining energies characterizing spin dependent couplings
between the band carriers and electrons localized in the open magnetic shells. 
Surprisingly, however, a series of recent experiments on (III,Mn)V DMS points to our 
limited understanding  of the $s$-$d$ exchange interaction in this important material family.\cite{Heim01,Wolo03,Myer05}
The determined $s$-$d$ exchange integral appears to have much smaller magnitude,\cite{Heim01,Wolo03}
and even opposite sign\cite{Myer05} to that expected according to the present knowledge
on the origin of the $s$-$d$ coupling in tetrahedrally coordinated DMS.

In this paper, we list first a number of obstacles making a quantitative determination
of the exchange integrals in III-V DMS difficult. We then analyze an additional ingredient
of these systems, namely the presence of Coulomb potentials centered on the magnetic ions as well
as on compensating donors. We evaluate electron wave function in the field of negatively
charge magnetic impurities and show that the Coulomb repulsion reduces the
apparent magnitude of the $s$-$d$ exchange integral. Importantly, the effect increases with lowering
magnetic ion concentration $x$, and becomes particularly significant in the experimentally
relevant range, $x \le 0.2$\%.\cite{Heim01,Wolo03,Myer05}
 
In the case of archetypical II-VI DMS such as (Cd,Mn)Te, the ferromagnetic exchange 
interaction between the conduction
band electrons and Mn spins is described by 
$N_0\alpha \approx 0.2$~eV, where $N_0$ is the cation concentration
and $\alpha$ is the $s$-$d$ exchange integral. This value of $N_0\alpha$ is about two times smaller 
than that describing the ferromagnetic exchange interaction between the $4s$ 
and $3d$ electrons in the free Mn$^{+1}$ ion.\cite{Diet94a} This reduction is caused by 
matrix polarizability
and the fact that not only cation but also anion $s$-type wave functions contribute 
to the Bloch amplitude of the conduction band electrons.  In the case
of the valence band holes, the exchange energy results from the symmetry-allowed $p$-$d$
hybridization, the typical value of the exchange energy being $|N_0\beta| \approx 1$~eV.
Within the molecular-field (MFA) and virtual crystal approximations (VCA), 
the exchange spin-splitting of the two-fold degenerate conduction and 
four-fold degenerate valence band is then, 
$s_z\alpha M/g\mu_B$ and $j_z\beta M/g\mu_B$, where 
$s_z = \pm 1/2$ and $j_z = \pm 1/2; \pm 3/2$, respectively, $M = M(T,H)$ is spin magnetization
of the substitutional magnetic ions characterized by the Land\'e factor~$g$.  The proportionality
between exchange splittings and independently measured magnetization has been demonstrated by a variety
of magnetooptical and magnetotransport experiments, and has made it possible to determine
accurately the values of $N_0\alpha$ and $N_0\beta$ for a number of systems.\cite{Furd88,Diet94,Mats02} 

However, the above simple scenario has been called into question in several important cases.
First, the orbital and carrier contribution to the measured $M$ has to be taken
into account.\cite{Kacm01,Diet01} Second, when the exchange energy $|N_0\beta|$ becomes
comparable to the valence band width, the MFA and VCA break down, particularly
in the range of small magnetic ion concentrations.\cite{Beno92,Twor94}
Third, the magnitude and sign of $\beta$ depend on the relative position of the
$p$ and $d$ states.\cite{Kacm01,Hart04} If, therefore, the charge
state and thus the energy of the relevant $d$ levels can be altered
by the position of the Fermi energy, the character of $p$-$d$ exchange will cease to be universal
in a given material but instead will depend on the doping type and magnitude. Fourth,
the intensity, and even the sign of the magnetic circular dichroism is strongly affected
by the Moss-Burstein effect. Accordingly, a simple relation 
between positions of the absorption edge for two circular light polarizations
and the splitting of the bands breaks down
in the presence of the delocalized or weakly localized carrier liquid.\cite{Szcz99,Diet01}  This
may account for the sign reversal of the apparent $\beta$ on going from
$n$-type to $p$-type (Ga,Mn)As.\cite{Hart04} 
Finally, spin-orbit interactions and $k\cdot p$ mixing between bands make spin-splitting away
from band extrema to be a complex non-linear function of $\alpha M$ and  $\beta M$ as well as
of relevant $k\cdot p$ parameters. This, in particular, has precluded a conclusive determination of the values
of the $sp$-$d$ exchange integrals for narrow-gap DMS of mercury and lead chalcogenides.\cite{Furd88,Diet94a} 
Such multi-band effects are especially important in quantum structures, where dimensional 
quantization enhances the kinetic energy of the carriers and, thus,
the effects of the $k\cdot p$ coupling. 
Indeed, an anomalous behavior of electron
spin-splitting in quantum wells of (Cd,Mn)Te and (Ga,Mn)As has been assigned 
to the $k\cdot p$ admixture of the valence band states to the electron wave function.\cite{Myer05,Merk99} 

Despite the difficulties in the precise determination of the exchange integrals, particularly
in quantum structures and systems containing carriers, a series of recent experiments
suggesting anomalous magnitude and sign of  $\alpha$ in (III,Mn)V DMS\cite{Heim01,Wolo03,Myer05} 
call for a detail consideration. In particular, Heimbrodt
{\em et al.}\cite{Heim01} detected spin-flip Raman scattering of conduction band electrons in
Ga$_{1-x}$Mn$_x$As, and evaluated  $N_0\alpha = 23$~meV for $x = 0.1$\%. Even a lower value
$|N_0\alpha = 14 \pm 4|$~meV was found by Wo{\l}o\'s {\em et al.},\cite{Wolo03} 
who analyzed the broadening by the electrons of the Mn spin resonance line in
$n$-Ga$_{1-x}$Mn$_x$N with $0.01\% \le x \le 0.2\%$. 
More recently, Myers {\em et al.}\cite{Myer05} examined
spin precession of the electrons in Ga$_{1-x}$Mn$_x$As quantum wells of the thickness
between 3 and 10~nm, and Mn content $x$ up to 0.03\%. As a result
of aforementioned admixture of the valence band states, the observed sign
of the exchange splitting is negative. The value $N_0\alpha = -90 \pm 30$~meV was determined
under a simplified assumption that the spin-splitting is 
proportional to magnetization, and
by extrapolating the resulting apparent exchange energy $N_0\alpha$ to the infinite quantum 
well width.\cite{Myer05} In contrast to the striking finding listed above, a large positive value $N_0\alpha 
\approx 0.5$~eV is consistent with intraband magnetoabsorption in $n$-In$_{1-x}$Mn$_x$As with 
a relatively high Mn content, $x \ge 2.5$\%.\cite{Zudo02}

We point here to an additional mechanism that may contribute to the anomalous
behavior of electron-spin splitting in III-V DMS. We note that 
the electric charge of the ${\rm Mn}^{2{+}}$ ion replacing e.g.\ a ${\rm Ga}^{3{+}}$ ion in the
lattice of a III-V compound (like $\rm GaN$ and $\rm GaAs$) is a
source of a repulsive electrostatic potential. Furthermore, the studied
samples are either $n$-type\cite{Wolo03} or at least highly compensated,
as evidenced by the presence of electron spin-flip Raman scattering\cite{Heim01}
and donor-related luminescence.\cite{Myer05} This indicates the
existence of attractive potentials associated with ionized {\em non-magnetic} donors. Thus,
the probability of finding a conduction band electron at the core
of the magnetic ion is reduced, and hence the \emph{apparent} value of the
exchange energy (the observed spin splitting) is diminished. It worth
noting that a possibility that the  Coulomb potentials could affect the apparent
value of the exchange integrals has already been mentioned in the context
of divalent Mn in GaN,\cite{Wolo03} and trivalent Fe in HgSe.\cite{Wila88}

\begin{figure}[htb]
\includegraphics[width=0.9\columnwidth]{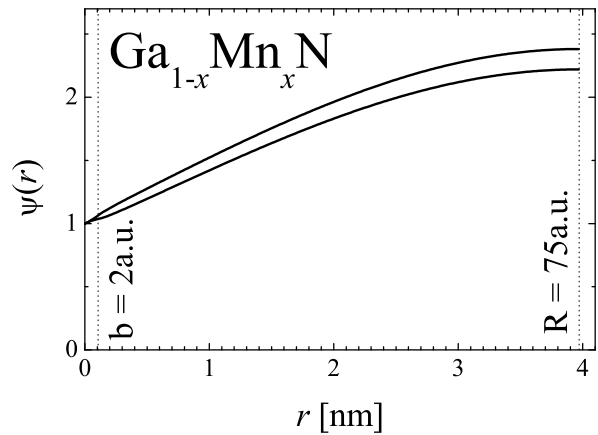}%
\caption[]{Wave functions of spin-up and spin-down carriers (the Coulomb term
  included) for ${\rm Ga}_{1-x}{\rm Mn}_x {\rm N}$ and $R = 75 \, \rm
  a.u.$ ($x = 0.0087 \%$).}
\label{psi75N}
\end{figure}

\begin{figure}[htb]
\includegraphics[width=0.9\columnwidth]{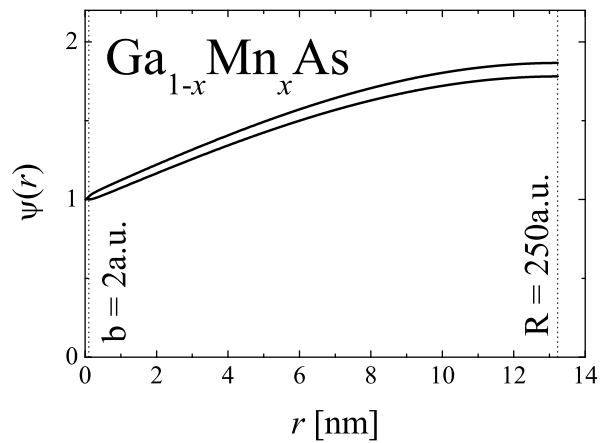}
\caption[]{Wave functions of spin-up and spin-down carriers (the Coulomb term
  included) for ${\rm Ga}_{1-x}{\rm Mn}_x {\rm As}$ and $R = 250 \,
  \rm a.u.$ ($x = 0.00047 \%$).}
\label{psi250As}
\end{figure}

To evaluate a lower limit of the effect we neglect the
presence of compensating donors and calculate the apparent $s$-$d$ exchange
integral $\alpha_{ap}$ for an electron subject to the repulsive potential generated
by the Mn acceptors. We follow a Wigner-Seitz-type approach put forward by one
of us and co-workers\cite{Beno92} to describe the interaction of the carrier spin with
the Mn ions in the case of the strong coupling limit, that is when the
depth of the local Mn potential is comparable to the carrier band width. It has
been found in the subsequent works\cite{Twor94} that the corrections to the
Wigner-Seitz approach caused by a random distribution of Mn ions are
quantitatively unimportant.

We consider a  Mn ion with the $5/2$ spin $\vec S_i$
located at $\vec R_i$, which interacts with
the carrier via the Heisenberg term $I(\vec r
- \vec R_i)\vec s \cdot \vec S_i $. 
The form of the function $I(\vec r - \vec R_i)$ makes
the interaction local: it vanishes outside the core of the Mn
ion. For simplicity, $I(\vec r - \vec R_i) = a \, \theta(b - |\vec r -
\vec R_i|)$. The exchange energy is then $\alpha = \int d^{3} \vec r
\, I(\vec r) = a \cdot \frac{4}{3} \pi b^3$. Moreover, in case of 
III-V compounds considered here, the impurity generates an electrostatic
potential. If screening by the electrons is present, as in case of $n$-${\rm
  Ga}_{1-x}{\rm Mn}_x{\rm N}$, this potential is
$e^2\exp(-\lambda r)/(4\pi\varepsilon_0 \varepsilon r)$, 
where $\varepsilon$ is the static dielectric constant, 
and the screening parameter $\lambda$ is given by
$\lambda^2 = e^2 \mathcal{N}(\mathcal{E}_F)/(\varepsilon_0 \varepsilon )
$, where $\mathcal{N}(\mathcal{E}_F) =
\frac{3}{2} n/k T_F$.\cite{Zima72}
For the ${\rm Ga}_{1-x}{\rm Mn}_x{\rm N}$ samples,\cite{Wolo03} $n
\approx 10^{19}\,{\rm cm}^{-3}$ corresponds to $T_F \approx 890 \, \rm
K$ ($\mathcal{E}_F \approx 0.12 \, \rm eV$), and therefore $1/\lambda
\approx 1.6 \, \rm nm $.  

\begin{figure}[tb]
\includegraphics[width=0.9\columnwidth]{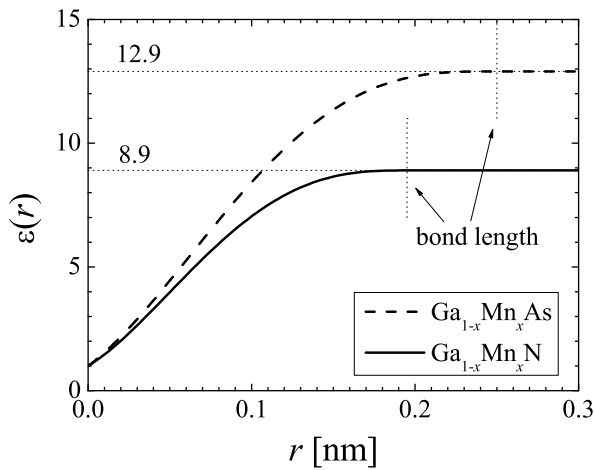}%
\caption[]{The assumed dependence $\varepsilon = \varepsilon(r)$.}
\label{epsiloncurve}
\end{figure}

In the spirit of the Wigner-Seitz approach we assume that the carrier
energy~$E$ and the envelope function~$\psi(r)$ are given by the ground
state $s$ solution of the one-band effective mass equation which
contains the potential $U(r)$ created by the magnetic ion located at
$r = 0$. The standard one-impurity boundary condition $\psi(r) \to 0$
for $r \to \infty$ is replaced by the matching condition $\psi'(r) =
0$ at $r = R$ to take into account the presence of other magnetic
ions. The value~$R$ is determined by the concentration of the magnetic
ions~$x$ according to the equation $(4 \pi R^3/3)^{-1} = N_0 x$. 
The exchange interaction is modelled by a square well
potential $U \theta(b-r)$ superimposed on the electrostatic potential
of an elementary charge located at $r = 0$. The potential $U = \pm \frac54 a$ is, of
course, different for spin-down and spin-up carriers.  

\begin{figure}[tb]
\includegraphics[width=0.9\columnwidth]{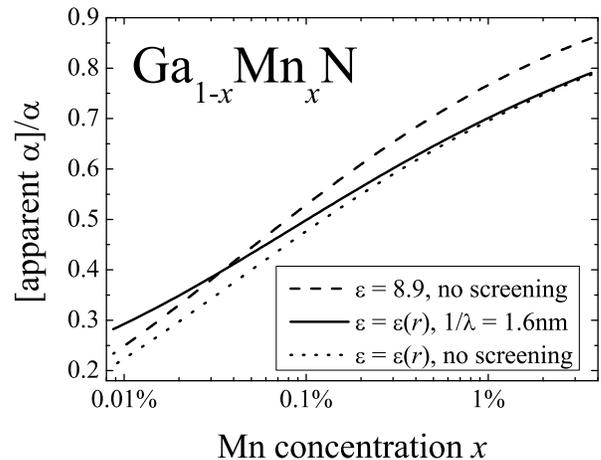}%
\caption[]{The dependence of the ratio of the apparent and bare exchange energies
$\alpha$ on~$x$ for for ${\rm Ga}_{1-x}{\rm Mn}_x {\rm N}$ and various models of screening.}
\label{xdepN}
\end{figure}

We first ignore free carrier screening, $\lambda \to 0$. The solution of
the time-independent Schr\"odinger equation for the conduction band
electron is then 
\begin{equation}
  \label{eqpsi1}
  \psi(r) = c_0 \exp(-\beta r) \Phi(1+\frac{A}{\beta}; 2; 2 \beta r) \equiv c_0 f
\end{equation}
for $0 < r < b$, and the following linear combination for $b < r < R$
\begin{eqnarray}
  \psi(r) & = & c_1 \exp(-\beta' r) \Psi(1+\frac{A}{\beta'}; 2; 2 \beta' r) + {} \nonumber
    \\
    & & {} + c_2 \exp(\beta' r) \Psi(1-\frac{A}{\beta'}; 2; -2 \beta'
    r) \nonumber \\ & \equiv & c_1 g + c_2 h, \label{eqpsi2}
\end{eqnarray}
where $A = e^2 m^{*}/(4\pi \varepsilon_0 \varepsilon \hbar^2)$,
$\beta = [2 m^{*}(U-E)]^{\frac12}/\hbar$, $\beta' = [2
m^{*}(-E)]^{\frac12}/\hbar$ (notice that changing the sign of $\beta$
leaves $\psi$ invariant, while changing the sign of $\beta'$
interchanges $c_1$ with $c_2$; also, $\Phi$ and $\Psi$ are not in
general linearly independent). We used the symbols $\Phi$, $\Psi$ for
the confluent hypergeometric functions ${}_{1}F_{1}(a; b; z)$, $U(a;
b; z)$.\cite{Slat60} The constants $c_0$, $c_1$, $c_2$ are determined
by the continuity conditions $\psi(b^{-}) = \psi(b^{+})$,
$\psi'(b^{-}) = \psi'(b^{+})$. Solving those two equations we obtain
an equation for~$E$,
\begin{equation}
  \label{eqdpsiReq0}
  \frac{w_{f, h}(b) g'(R) - w_{f, g}(b) h'(R)}{w_{g, h}(b)} = 0,
\end{equation}
where by $w_{f, g}$ we denoted the Wronskian $f g' - f' g$.
In the following, $\psi(r)$ is normalized as $\psi(0) = c_0 = 1$.

We assume the following parameters for ${\rm Ga}_{1-x}{\rm Mn}_x {\rm
  N}$: $m^{*} = 0.22 \, m_e$, $N_0 = 4.38\cdot10^{22} \, {\rm cm}^{-3}
= 0.006495 \, \rm a.u.$, $\varepsilon = 8.9$; and the following for
${\rm Ga}_{1-x}{\rm Mn}_x {\rm As}$: $m^{*} = 0.067 \, m_e$, $N_0 =
2.21\cdot10^{22} \, {\rm cm}^{-3} = 0.003281 \, \rm a.u.$,
$\varepsilon = 12.9$. In the experiments, samples were used with $0.01\% \le x \le 0.2\%$ of
$\rm Mn$ in $\rm GaN$,\cite{Wolo03} and with $0.0006\% \le x \le 0.03\%$
of $\rm Mn$ in $\rm GaAs$.\cite{Myer05} Those concentrations
correspond to $R$ up to about $75 \, \rm a.u.$ for $\rm GaN$ and up to
about $250 \, \rm a.u.$ for $\rm GaAs$.

To visualize the effect of the Coulomb term in the
Mn potential, we have calculated the energies and wave functions
including and disregarding the additional Coulomb term for both 
GaN ($b = 2 \, {\rm a.u.} \approx 0.1 \, \rm nm$, $a = 0.0371 \, \rm
a.u. = 1.0 \rm\, eV$) and $\rm GaAs$ ($b = 2 \, \rm a.u. \approx 0.1
\, \rm nm$, $a = 0.0735 \, {\rm a.u.} = 2.0 \, \rm eV$). 
These parameters correspond to $N_0 \alpha = 0.22 \,
\rm eV$, a value for CdS.\cite{Beno92} We have found that
when calculating $\alpha_{ap}/\alpha$, the
details of the exchange potential (like the values of~$b$ and $\alpha$
within the expected range) are not 
quantitatively important. 

In order to take into account the fact that the core and lattice
polarizability decrease at small distances, $\varepsilon \rightarrow 1$ for $r \rightarrow 0$,
we interpolate $\varepsilon(r)$ between $\varepsilon(0) = 1$ and the macroscopic value
attained at a distance of the bond length. The 
assumed dependence, presented in Fig.~\ref{epsiloncurve}, is similar to
that of the Thomas-Fermi model.\cite{Rest77}
When  $\varepsilon = \varepsilon(r)$ and/or free carrier screening is included, 
we find the solution $\psi(r)$ of the Schr\"odinger equation for
the given potential $U(r)$ numerically, as Eqs.~(\ref{eqpsi1}) 
and~(\ref{eqpsi2}) are only valid for the Coulomb
potential. Then, the spin splitting for a given value of~$x$ (or for
the corresponding~$R$) is evaluated as the difference of the
energy~$E$ calculated for the spin-up and spin-down carriers from the
equation $\psi'(R) = 0$. Here, $\psi(r)$ is the numerical solution of
the Schr\"odinger equation with the potential that is different for
spin-up and spin-down carriers.

\begin{figure}[tb]
\includegraphics[width=0.9\columnwidth]{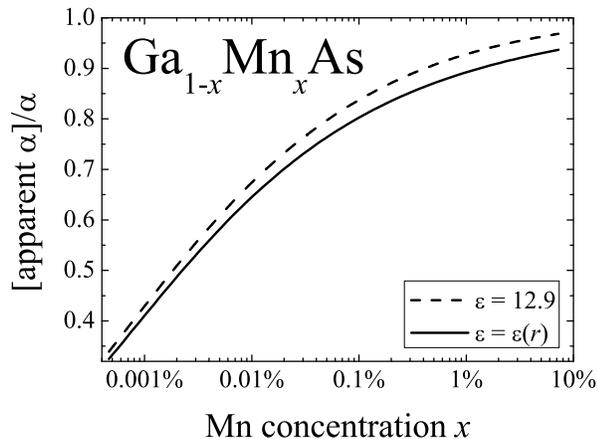}%
\caption[]{The dependence of the ratio of the apparent and bare exchange energies
$\alpha$ on~$x$ for ${\rm Ga}_{1-x}{\rm Mn}_x {\rm As}$.}
\label{xdepAs}
\end{figure}

The results of our calculations of $\alpha_{ap}/\alpha$ as a function
of the Mn ion concentration $x$ are presented in Fig.~\ref{xdepN}
(${\rm Ga}_{1-x}{\rm Mn}_x {\rm N}$) and in Fig.~\ref{xdepAs} (${\rm
Ga}_{1-x}{\rm Mn}_x {\rm As}$). Independently of assumptions
concerning screening, in both materials  $\alpha_{ap}/\alpha$ diminishes
significantly when $x$ decreases, up to factor of three in the experimentally relevant
range of~$x$. However, this reduction of $\alpha_{ap}/\alpha$ is still smaller
than that seen experimentally,\cite{Heim01,Wolo03} presumably because of an additional 
effect coming from the presence of attractive potentials brought about by
compensating non-magnetic donors.

In summary, we have enlisted a number of effects that renders an accurate
experimental determination of the $sp$-$d$ exchange integrals difficult, particularly
in cases when both $p$-like and $s$-like states contribute to the carrier
wave function. The interaction of conduction band electrons with the
magnetic ions in (${\rm Ga}_{1-x}{\rm Mn}_x {\rm N}$, ${\rm Ga}_{1-x}{\rm
Mn}_x {\rm As}$) has been
considered quantitatively taking into account the electrostatic potential created by the
magnetic ion. A substantial reduction in the magnitude
of the apparent exchange energy has been found at low Mn concentrations,
and interpreted as
coming from the decrease of the carrier probability density at the
core of the magnetic ion caused by the electrostatic repulsion. It has
been suggested that this effect, enhanced by an attractive potential of
compensating donors, accounts for abnormally small values of the
exchange spin splitting observed experimentally in III-V DMS containing 
a minute amount of Mn.\cite{Heim01,Wolo03,Myer05} In view 
of our findings, the presence of electrostatic potentials associated with magnetic
ions makes that
the apparent exchange energies should not be viewed as universal but rather dependent on the 
content of the magnetic constituent and compensating donors.

\end{document}